**Room-temperature superconductivity in ultra-thin carbon nanotube zeolite composites: a conventional or unconventional superconductor?**

[1]Chi Ho Wong, [2]Rolf Lortz

[1]Division of Science, Engineering and Health Studies, School of Professional Education and Executive Development, The Hong Kong Polytechnic University, Hong Kong, China

[2]Department of Physics, The Hong Kong University of Science and Technology, Hong Kong, China

**Abstract:**

The recent report of signs of room-temperature superconductivity in ultrathin single-walled carbon nanotubes (CNT) of types (2,1) and (3,0) holds significant promise for energy applications due to their ability to conduct current without dissipation. However, the McMillan Tc formula fails to calculate their superconducting transition temperatures (Tc) accurately, which raises an important question: what is the pairing mechanism driving their room-temperature superconductivity? To explore this further, we first investigate whether the strong curvature of ultrathin CNT leads to exotic phenomena in unconventional superconductors. If no evidence of these exotic characteristics is found and the McMillan formalism indicates that it is not a BCS-type superconductor, could we be observing a new class of unconventional superconductivity that functions independently of phonons and typical exotic features? In this paper, we demonstrate that factors such as the chiral angle of CNT, boron dopants and lattice regularity can be used to tune the theoretical $T_c$ to experimental values. Our finding suggests that combining CNT with a harder substrate could be vital for further enhancing Tc while minimizing lattice distortion under doping. We propose a reconsideration of the common belief regarding whether the McMillan and BCS Tc formulas are adequate for classifying materials as BCS or non-BCS superconductors.

**1. Introduction:**

The journey of superconductivity began with Kamerlingh Onnes's discovery of zero resistance in mercury in 1911, marking a pivotal milestone in physics [1]. Several decades later, in 1987, Paul Chu aimed for higher superconducting transition temperatures (Tc) of 90 K, achieving this with $YBa_2Cu_3O_7$ (YBCO) [2]. Notably, other cuprate family members can have their Tc values tuned to approximately 160 K through pressure or doping [3]. Two decades later, the landscape shifted with the discovery of iron-based superconductors (IBSC) by Hideo Hosono in 2008 parallelly [4]. While bulk IBSC materials demonstrate superconductivity at around 55 K [5], low-dimensional structures like FeSe can achieve Tc values near 100 K via the proximity effect [6]. Nichelates can also show superconductivity around 80K in recent years. The primary motivation behind all these efforts is the pursuit of room-temperature superconductivity, which has the potential to revolutionize technologies by enabling energy transmission without dissipation.

Despite the promising high Tc values in unconventional superconductors, the lack of a robust theoretical model complicates the path toward the development in room temperature superconductors. The Bardeen-Cooper-Schrieffer (BCS) model effectively describes conventional superconductors, emphasizing electron-phonon coupling in the absence of magnetism [7]. Fine-tuning Tc calculations for BCS superconductors can be accomplished using the McMillan Tc formula [8]. However, the emergence of unconventional superconductors has complicated this narrative. These unconventional materials often exhibit exotic features such as spin-orbit coupling, strong correlation of electrons, antiferromagnetism, and anisotropic Fermi surface, along with charge density waves and/or spin density waves, etc, which makes the modelling complicated [9-14]. In 2024, we proposed a formalism to calculate the Tc values of a number of iron-based and cuprate superconductors under various pressures and doping levels, achieving results very close to experimental Tc [15-19]. Based on the WL formalism, the strength of the differential phonon could be a crucial factor for deactivating the isotropic effect in IBSC and reactivating it in YBCO [15-19]. However, the formalism still falls short of capturing the complexities necessary for deriving a universal Tc formula, as it only considers the synergistic effects of higher-order antiferromagnetism, spin-orbit coupling, anisotropic Fermi surface, refinements in the electron concentration based on ARPES data, electron-differential phonon coupling, unusual screening effect under charge density waves, and/or spin density waves, among others [15-19].

Since the last decade, compressing hydride gases has shown promise in approaching room-temperature superconductivity based on the BCS mechanism. Although high Tc of over 200K has been achieved in the compressed hydrides [20], the extreme pressures required render this approach impractical for everyday applications. Our theoretical quest for high-temperature superconductivity in one-dimensional carbon structures has been called 2017 [21]. The main challenge was that the BCS or McMillan Tc formula requires a constant electronic density of states (DOS) around the Fermi level [8]. However, one-dimensional materials typically exhibit a Van Hove singularity in their electronic states. Hence, neither the BCS nor McMillan Tc formula can accurately calculate the Tc of one-dimensional carbon structures [21]. In view of this discrepancy, C.H. Wong et al. developed a scale factor approach for extreme conditions, specifically when the Debye temperature is significantly higher than Tc [21]. In this scenario, it can be derived that by comparing two materials, the ratio of their Tc values is approximately equal to the ratio of their electron-phonon coupling constants [21,22]. After validating this statement using a test dataset of well-known bulk BCS superconductors (e.g., Pb, Al, Sn, etc.) [21], C.H. Wong et al proceeded to calculate the Tc of 4Å single-walled carbon nanotubes (SWCNT) with a good agreement with experimental Tc [21-23]. Then we have proposed that the optimized kink structure in a carbon-ring structure under cumulative phase conditions can reach Tc above 100 K [21].

This year, through a collaborative effort led by Sheng et al., ultrathin boron-doped carbon nanotubes with chiral angles of (3,0) and (2,1) indicate an impressive ambient pressure experimental Tc of 230-250 K [24]. Applying slight pressure can boost the Tc to an impressive 400 K [24]. However, the underlying mechanisms driving this room-temperature superconductivity remain unclear. We aim to investigate any exotic features typical of unconventional superconductors present in these ultrathin CNT. If such exotic features are absent, and the Tc value calculated using the McMillan Tc formula indicates it should not be BCS-type, could there be a brand-new formation of unconventional superconductivity without the exotic effects and electron-phonon coupling? Our insight will be provided in this work.

## 2. Computational Methods

We revisit the scale factor approach briefly [21,22]. The pairing Hamiltonian $H_{pair} = \sum_{k\sigma} E_k n_{k\sigma} + \sum_{kl} V_{kl} c_{k\uparrow}^* c_{k\downarrow}^* c_{l\uparrow} c_{l\downarrow}$ consists of the single-particle energy $E_k$ measured against the Fermi energy, along with an interaction term $V_{kl}$ that accounts for the scattering of a particle from one state to another, specifically from a state with $(l\uparrow, -l\downarrow)$ to a state with $(k\uparrow, -k\downarrow)$. The operators $c_{k\uparrow}^*$ and $c_{k\downarrow}^*$ are creation operators for spin-up and spin-down particles, respectively, while $n_{k\sigma}$ serves as the particle number operator. The spin index is denoted by $\sigma$. The ground state of the BCS wavefunction is $|\psi_G\rangle = \prod_{k=k_1....k_M} \left(u_k + v_k c_{k\uparrow}^* c_{k\downarrow}^*\right)|\varphi_0\rangle$, where $|\varphi_0\rangle$ is the vacuum state in the absence of particles and $|u_k|^2$ is the unoccupied probability due to $|u_k|^2 + |v_k|^2 = 1$. The energy gap $\Delta$ is independent of *k*. Based on an isotropic k space, the gap is $\Delta = \Delta_k = -\sum_l V_{kl} u_l v_l$ [21,22], which becomes $-\Delta \sum_k u_k v_k = \sum_{kl} V_{kl} u_k v_k u_l v_l$ by multiplying $-\sum_k u_k v_k$ to both sides of the formula. This can be further simplified into $\Delta = \frac{H_{e-ph}}{-\sum_k u_k v_k}$ [21,22]. Given that the ratio of $-\sum_k u_k v_k$ between two superconductors P and Q equals to ~1, the scale factor approach in the context of $\frac{\Delta_P}{\Delta_Q} \sim \frac{H_{e-ph(P)}}{H_{e-ph(Q)}}$ is fulfilled [21,22]. In other words, the $T_c$ of the material *P* can be predicted if the $T_c$ of the material *Q* and the ratio of electron-phonon coupling are known.

However, a transfer function is needed due to the dependence between $u_l v_l$ and $\Delta$ [21,22]. Based on the BCS theory, the energy gap can be derived as $\Delta_k = -\frac{1}{2}\sum V_{kl} \frac{\Delta_l}{(\Delta_l^2 + E_k^2)^{0.5}}$ [21,22]. To obtain the transfer function, we first take the electrons at the Fermi level into account (or $E_k = 0$) and then set a trial energy gap $\Delta^T$. Eventually, the $\Delta^T$ is proportional to the interaction term $V_{kl}$ directly. The transfer function, i.e. $u_l^T v_l^T$ as a function of electron energy, is used to correct the energy gap, i.e. $\Delta_k^{corrected} = -\sum_{kl} V_{kl} u_l^T v_l^T$ [21,22]. To construct a reasonable transfer function, the Debye temperature should be much higher than the Tc. If not, the BCS occupational fraction will not quickly approach zero as the electron energy increases. Given that $T_D$ is Debye temperature, $\lambda$ represents electron phonon coupling, $\mu$ is Coulomb pseudopotential and $f(T_D)$ is the ratio of transfer function, we define $T_c^P = T_c^Q \left| \frac{\lambda_P - \mu_P}{\lambda_Q - \mu_Q} \frac{1}{f(T_D)} \right|$ [21,22]. Our previous studies on the scale factor approach indicate that a Debye temperature several times greater than Tc would introduce at most a 1% error in the Tc calculation, resulting in f($T_D$) values between approximately 1.001 and 1.01 [21]. Additionally, the value of μ is much smaller than λ [21]. Then we set f($T_D$) = 1 and approximate μ as 0. Unless otherwise specified, (5,0) CNT will serve as the reference material for the scale factor approach. We will calculate the Tc for (3,0)CNT, (2,1)CNT, porous (2,1)CNT, boron-doped (2,1)CNT, and porous boron-doped (2,1)CNT accordingly.

The GGA-PBE functional in CASTEP is employed (unless specified otherwise) [25]. The electronic band diagram and electronic density of states [25] are calculated where the energy cut-off point and tolerance are set at 240 eV and 2μeV/atom, respectively. An ultrasoft pseudopotential is utilized. Density mixing has been selected for the setup of the electronic minimizer. A wall-to-wall distance of 5.5Å, resulting in a central-line separation of ~8Å laterally is set. Due to the circular shape in CNT, the curvature effect of electrostatics is approximately modified by $Z_{effective}$ [21], where $Z_{effective} = Z \frac{\sum_{i=0}^{\infty} U_P(i)}{\sum_{i=0}^{\infty} U_Q(i)}$ can be determined by evaluating the ratio of the lattice potentials as a function of lattice grid (i) between two reference materials with distinct curvatures, and Z is the atomic number of carbon [21]. Geometric optimization is performed. Phonon computations are based on the finite displacement algorithm [25]. Phonon softening along longitudinal, tangential and chiral axis are considered by vector components in harmonic approximation (See supplementary materials). The selection of spin-orbital coupling and magnetism are activated for relevant analysis.

## 3. Results and Discussion

### 3.1. The Quest for the Exotic Features:

To analyze the origin of room-temperature superconductivity in ultrathin CNT, we employ a spin-unrestricted ab-initio setup to confirm that both (3,0) CNT and (2,1) CNT are non-magnetic, irrespective of the presence of dopants or vacancies. In this non-magnetic context, the formation of a spin density wave is not feasible [14-19]. Across the maxima and minima of the spin density wave, a change in the microscopic magnetic field is triggered, which presumably induces periodic charge modulation based on Maxwell's equation [14-19]. This can be one reason for the formation of a charge density wave [14-19]. In the absence of a spin density wave, we also observe no charge density wave in the ab-initio data ]14-19]. Without complex electron correlations on the Fermi surface, we consistently find an isotropic Fermi surface in both (3,0) CNT and (2,1)CNT, regardless of dopants or vacancies.

### 3.2. A Test of the McMillan Tc Formula

We compare the average electronic DOS($E_F$) of the (5,0)CNT, (3,0)CNT, and (2,1)CNT in Figure 1a. While the experimental Tc of (5,0)CNT is 7.6 K [23], the higher DOS observed in (3,0)CNT, (2,1)CNT and 3.5% B-doped CNT could potentially lead to a higher Tc if the pairing mechanism is predominantly of the BCS type. Our ab-initio data shows that the spin-orbit coupling these CNT can be ignorable. Based on the McMillan Tc formula, we can maximize the Tc of the B-doped (2,1)CNT by adjusting the Coulomb pseudopotential to zero. With a Debye temperature of 1500 K and an electron-phonon coupling constant $\lambda$ of ~0.2, the calculated Tc (McMillan) is only ~2K. At first glance, the room-temperature superconductivity of the ultrathin CNT does not appear to follow the BCS mechanism.

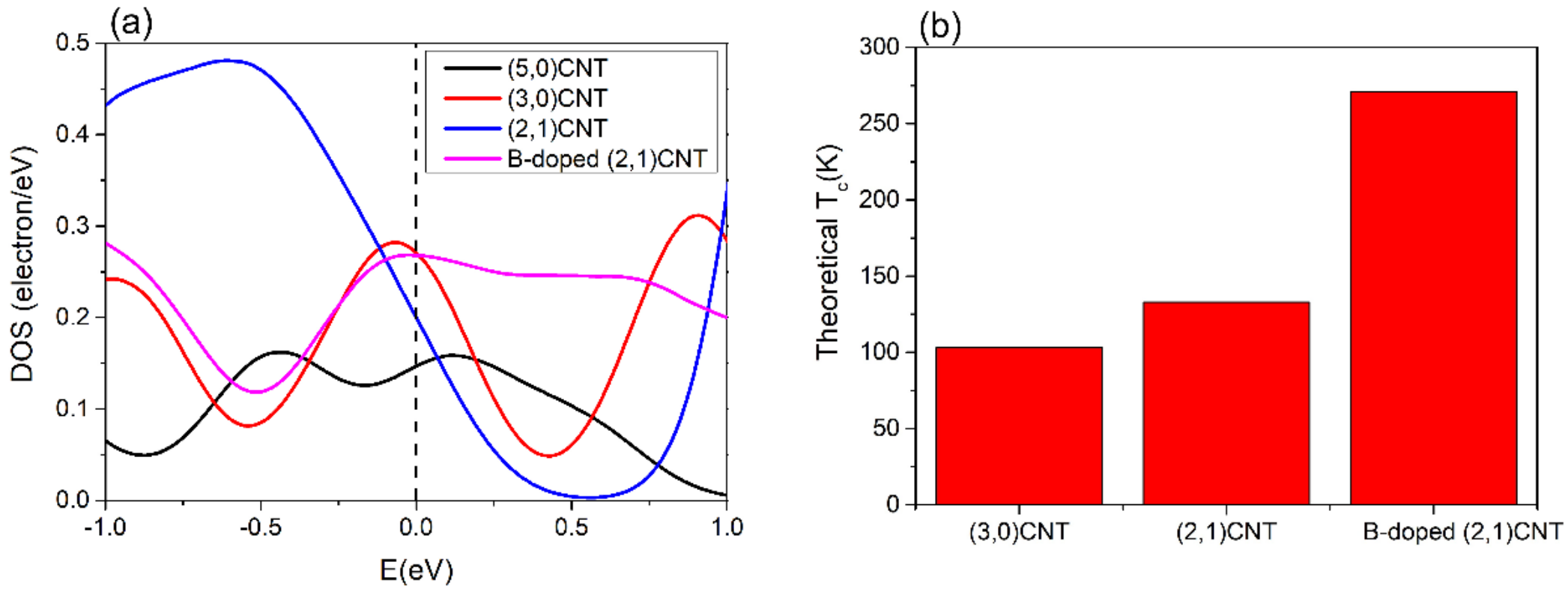

Figure 1: (a) Electronic DOS per atom in (5,0)CNT, (3,0)CNT, (2,1)CNT and 3.5% B-doped (2,1)CNT are shown. (b) Our calculated Tc values of (3,0)CNT, (2,1)CNT and 3.5% B-doped (2,1)CNT based on the scale factor approach. All the CNT structures are further relaxed under geometric optimization under a lateral distance of ~8Å.

A room-temperature superconductor made of carbon, despite lacking exotic features and exhibiting a low $T_{c_McMillan}$, raises a question about what drives the strong pairing interaction among electrons. Two possible explanations could be: the ultrathin CNT may possess previously unrecognized exotic features that have not been reported in the search for iron-based and cuprate superconductors, suggesting the need to define a new class of unconventional superconductors. Alternatively, the McMillan $T_c$ formula may not be applicable in this scenario. This can be a debatable situation. The former explanation spans a timeline dating back to 1986, with numerous relevant works published over the years. Considering the latter as a potential explanation seems promising, we use the scale factor approach to calculate the Tc for these CNT configurations at a reasonable level. This method is specifically designed to address the Van Hove singularity in the DOS [26] within nanowire structures with high Debye temperatures.

### 3.3. Why does (2,1) CNT have a higher Tc despite a lower DOS($E_F$) than (3,0) CNT?

For the (3,0)CNT, we obtain a ratio of effective atomic spring constant ($K_{resultant}$ ratio) = 0.74 (see supplementary material). This indicates strong phonon softening due to curvature when compared to (5,0) case. In (3,0) CNT, the DOS($E_F$) rises by 1.78 times in comparison to (5,0)CNT. With the revision of curvature effect of electrostatics, $\langle g_{kk}^2 \rangle$ is 5.64, where $g_{kk}$ is the electron-phonon scattering matrix. Substituting these parameters into the scale-factor approach [21-22], the theoretical Tc of (3,0)CNT is 105K in Figure 1b. (see supplementary material). On the other hand, for the (2,1)CNT, we find $K_{resultant}$ ratio = 0.34 which indicates a stronger phonon softening. In (2,1) CNT, the $\langle g_{kk}^2 \rangle$ increases by a factor of 3.76, while the DOS($E_F$) increases by 1.34 times in comparison to (5,0) CNT. Using the scale-factor approach, the theoretical Tc of (2,1) CNT is estimated to be 137 K (refer to supplementary material) in Figure 1b. Despite the lower gains in $\langle g_{kk}^2 \rangle$ and DOS($E_F$) in (2,1)CNT compared to (3,0)CNT, the calculated Tc of (2,1)CNT is higher. This is because the tube curvature not only softens the phonon along zigzag path but also the armchair path (see Figure S1d in the supplementary materials), further enhancing phonon softening. Otherwise, the calculated Tc of (2,1)CNT is only 52K while considering the phonon softening along the zigzag path alone.

### 3.4. Van Hove Singularity in DOS($E_F$)

In our 2017 publication, we reported that a kink-structured carbon ring could achieve a Tc of 104 K [21]. If we consider CNT as a series of these kink-structured carbon rings arranged along the longitudinal axis, their resulting Tc values are indeed comparable. However, we did not explore the tuning of the DOS($E_F$) towards the Van Hove singularity 8 years ago. Figure 1a indicates that the DOS of (2,1) CNT has a peak at -0.55 eV with a negative slope. This suggests that by reducing pressure or introducing a light atom with an acceptor nature to shift the Fermi level further negative, the Tc could be significantly enhanced, where the manually shifted DOS at the point of Van Hove singularity exhibits an impressive two-fold increase.

Encouraged by the theoretical Tc exceeding 130K, doping with boron presents a promising option to increase the Tc. Being lighter than carbon, substituting a carbon atom with boron reduces internal pressure. Additionally, since boron is a p-type dopant, it helps lower the Fermi level, where the B-doped CNT under geometric optimization can reach a theoretical Tc of 277K in Figure 1b (see supplementary materials), close to the experimental Tc of 285 K. Doping boron atoms in (2,1)CNT reduces the internal pressure while increasing the average diameter of CNT by approximately 10%.

Figure 2a compares the increase in the DOS($E_F$) in (2,1)CNT after doping boron with and without lattice distortion. By replacing C atoms with B atoms while keeping the atomic coordinates constant (i.e fixed shape), we can isolate the effect on DOS($E_F$) without the influence of lattice distortion. This approach allows us to clearly observe the electronic characteristics of the 3.5% B-doped (2.1)CNT (fixed shape), where a two-fold increase in DOS($E_F$) is observed and a complete vanishing of spin-orbit coupling is also noted when boron is doped. The pairing interaction in the 3.5% B-doped (2.1)CNT, which maintains a fixed shape, gives a theoretical Tc of 438K. When comparing the theoretical Tc of the 3.5% B-doped (2.1)CNT in fixed shape to that of the relaxed shape, it becomes clear that maintaining structural integrity is crucial for enhancing Tc. If lattice distortion following doping can be minimized through the application of small pressure or by using a harder substrate, the theoretical Tc could reach as high as at least 438 K. In the experiment led by Lortz's team [24], the experimental Tc of the B-doped CNT reaches above 400K under a small pressure of 0.002GPa. The application of 0.002 GPa does not alter the lattice constant a lot, based on the large elastic modulus of CNT [27]. It is believed that the effect of 0.002 GPa in the experiment likely suppresses lattice distortion when boron is doped.

### 3.5. Sample Quality

Figure 2b illustrates how poor sample quality can undermine the fabrication of a room temperature superconductor. The (2,1)CNT has a theoretical Tc of 137 K; However, introducing just 3.5% vacancies reduces the theoretical Tc by nearly 10 times to 16.4 K, accompanied by a noticeable decrease in DOS($E_F$) (black vs. red curve, Figure 2b). Furthermore, adding 3.5% vacancies to the 3.5% B-doped (2,1)CNT lowers Tc by a factor of 2.71, from 271 K to 100 K, which is associated with a significant drop in DOS($E_F$) (green vs. blue curve, Figure 2b). This highlights the critical importance of quality control in sample fabrication. Notably, B-doped (2,1) CNT shows greater strength at the same level of porosity compared to its undoped counterpart. Therefore, observing the superconductivity in a pristine 2Å CNT at ~130K relies heavily on maintaining high sample quality. The Debye temperature of the CNT in this work is set to 1500 K [21,28], where the Debye temperature is higher than the Tc 4-5 times can satisfy that the ratio of the transfer function is very close to 1, with an error of Tc less than approximately 1 percent [21-22].

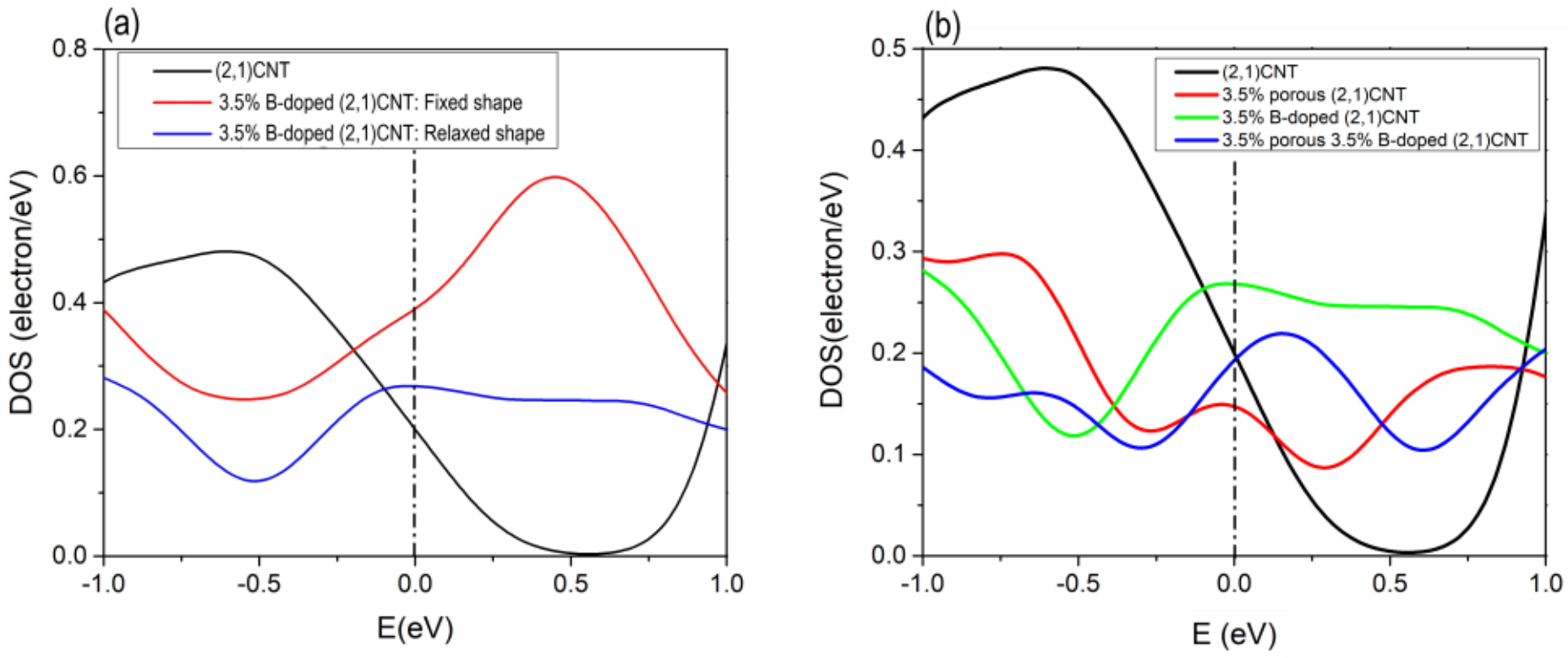

Figure 2: (a) The electronic density of states (DOS) of (2,1)CNT varies significantly with boron doping and lattice distortion. The 3.5% B-doped (2,1)CNT (fixed shape) refers to the scenario where B atoms are replaced without geometric optimization, while the 3.5% B-doped (2,1)CNT (relaxed shape) refers to the case where B atoms are replaced with geometric optimization in the full lattice structure. (b) The impact of porosity on the DOS in both undoped and 3.5% B-doped (2,1)CNT is illustrated.

### 3.6. Our Scientific Insights

Our analysis reveals several key insights into the pursuit of room-temperature superconductors. Worldwide efforts have focused on enhancing the Tc of unconventional superconductors, leveraging their inherently high starting Tc. However, the absence of a universal Tc formula complicates these searches. In contrast, we have demonstrated that the BCS mechanism can indeed facilitate room-temperature superconductivity in solid materials at ambient pressure. This finding eliminates the reliance on extreme pressures, as required by compressed hydride gases [20], and opens new avenues for transmitting dissipationless currents. Our theoretical predictions and experimental results suggest that modifying BCS solid materials at ambient pressure is a promising approach to design room-temperature superconductors. Our theoretical calculation serves as a foundation to prevent the misclassification of ultrathin CNTs as a distinct class of unconventional superconductors due to a good alignment between the experimental and calculated Tc.

In addition, we have reaffirmed the significance of the scale factor approach for calculating the Tc of ultrathin CNT [21,22]. This method effectively addresses the Van Hove singularity issue [26] encountered when applying the McMillan Tc formula, providing a more accurate framework for understanding superconductivity in these 1D materials. We do not underrate the significance of the McMillan Tc formula in one-dimensional contexts; rather, we find that under specific conditions: namely, 1D limit, Van Hove singularity and high Debye temperatures, the scale factor approach should provide a more precise Tc calculation.

A good alignment between theoretical and experimental Tc in these CNT systems suggests that, since the discovery of superconducting carbon nanotubes in 2001 [29], the accurate calibration of DFT functional regarding phonon softening for calculating the Tc of superconducting CNT remains an open question. This issue may lead to discrepancies between $T_{c_McMillan}$ and experimental results [21,22]. However, our vector analysis within the harmonic approximation [21,22] can provide a reasonable estimate of the strength of phonon softening for the superconductivity in CNT. We do not discredit ab-initio software in this ultrathin CNT context. Instead, it is a matter of time before a calibrated ab-initio setup tailored for the superconductivity in CNT is established. Although efforts to search for such ab-initio setups have been ongoing since 2001, progress is still being made. Once a calibrated ab-initio framework is identified for the superconductivity in CNT, it should provide more accurate Tc calculations than those derived from harmonic approximations. However, in the absence of this information, using a vector analysis of harmonic approximation remains the best alternative. Currently, there is no evidence that any exotic features in unconventional superconductors exist in ultrathin CNT. Therefore, we propose that the BCS mechanism is sufficiently strong to elevate the Tc of ultrathin CNT above room temperature.

**4. Conclusion:**

Our study demonstrates that room-temperature superconductivity at ambient pressure can be achieved in solid materials using the conventional BCS mechanism. We also show how the scale factor approach effectively addresses the superconductivity of 1D materials. Notably, the DOS at the Fermi level of (2,1)CNT is lower than that of (3,0)CNT; however, the Tc of (2,1)CNT is higher because (3,0)CNT primarily exhibits phonon softening along the zigzag path, while (2,1)CNT experiences phonon softening along both the armchair and zigzag paths. Furthermore, we propose that controlling lattice distortion in CNTs is crucial for maintaining the Tc value far above room-temperature. Based on the good alignment between the theoretical and experimental Tc, we raise an open question whether the McMillan Tc formula is sufficient to justify an ultrathin 1D material following BCS mechanism or not.

**Acknowledgement:**

We acknowledge the Department of Industrial and Systems Engineering at The Hong Kong Polytechnic University to provide simulation support.

**Conflict of Interest**

The authors declare no conflict of interest

**Data Availability Statement**

The data is sharable upon reasonable request.

**Supplementary materials**

Section 1: Vector analysis of harmonic approximation.

- Let's find out the average spring constant of the repeatable pattern of covalent bonds between atoms $C_1$ and $C_2$, $C_2$ and $C_3$, $C_3$ and $C_4$, and $C_4$ and $C_5$ along the longitudinal Z axis in Figure S1a [21,22]. Three possible combinations are listed in Figure S2.

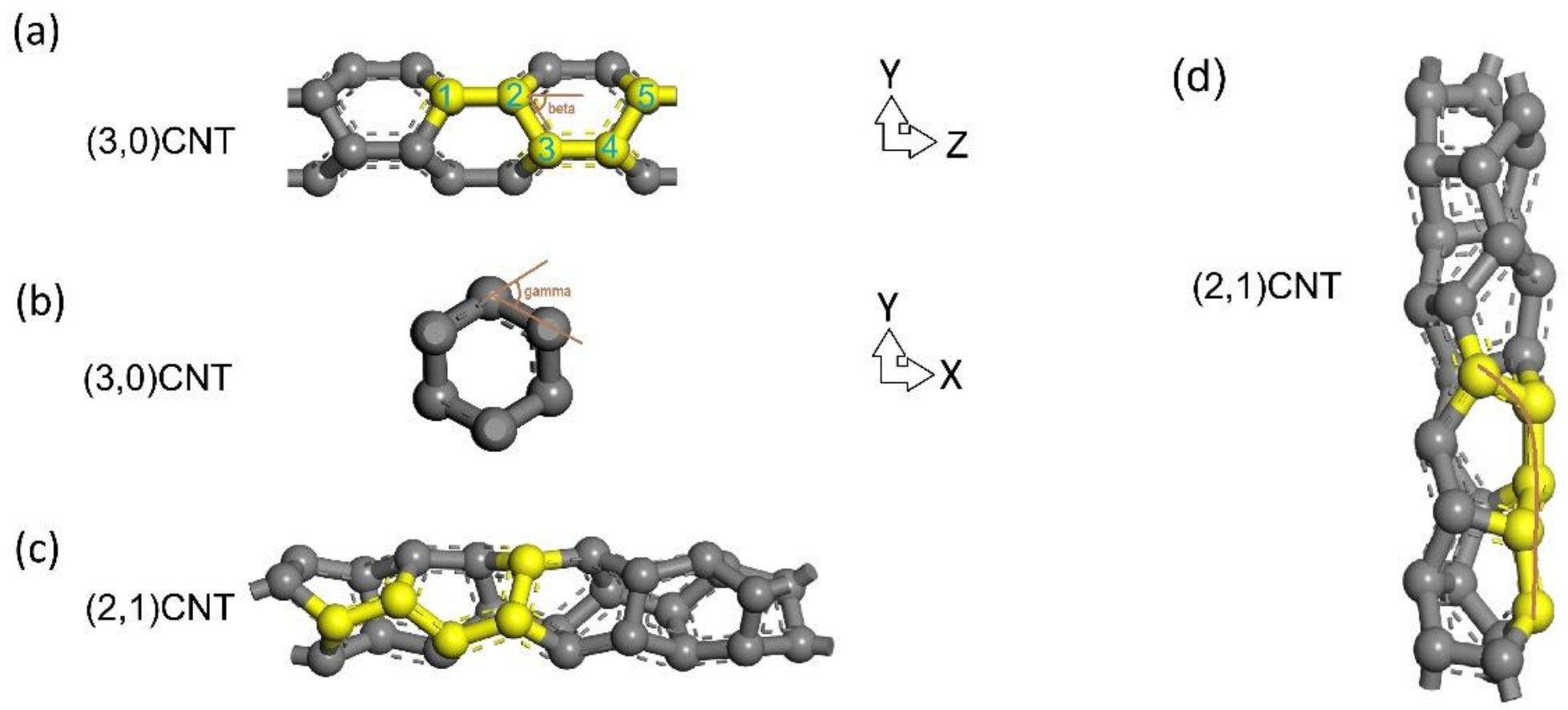

Figure S1: (a) Side view of the (3,0) CNT repeated unit in the YZ plane. Four covalent bonds are considered between atoms $C_1$ and $C_2$, $C_2$ and $C_3$, $C_3$ and $C_4$, and $C_4$ and $C_5$, all under translational symmetry. (b) Side view of the (3,0) CNT repeated unit in the XY plane. A covalent bond between two nearest neighbors is considered under rotational symmetry. (c) A convenient side view of the (2,1) CNT repeated unit, where the four covalent bonds in Figure S1(a) are highlighted correspondingly. (d) Another convenient view of the (2,1) CNT repeated unit. Unlike (3,0)CNT, we observe that the five carbon atoms (represented by yellow spheres) produce a gradual change in the tube curvature, i.e. $C_1$ and $C_5$ are not colinear.

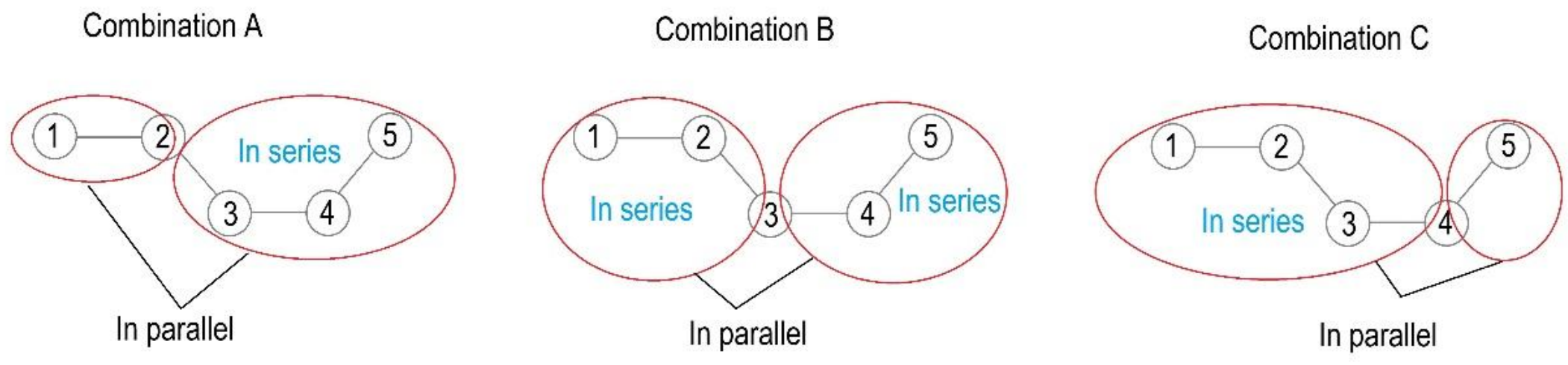

Figure S2: Three possible combinations for the vector analysis of harmonic approximation for the structure shown Figure S1a

Combination A:

Group bond $K(C_2\text{-}C_3)$, bond $K(C_3\text{-}C_4)$, bond $K(C_4\text{-}C_5)$ in series, then Bond $K(C_1\text{-}C_2)$ is in parallel with $K_{series}(C_2\text{-}C_3,C_3\text{-}C_4,C_4\text{-}C_5)$

$$K_{series}(C_2\text{-}C_3,C_3\text{-}C_4,C_4\text{-}C_5) = [(K(C_2\text{-}C_3)\cos(beta))^{-1} + K(C_3\text{-}C_4)^{-1} + (K(C_4\text{-}C_5)\cos(beta))^{-1}]^{-1}$$

$K(C_1\text{-}C_2)$ in parallel with $K_{series}(C_2\text{-}C_3, C_3\text{-}C_4, C_4\text{-}C_5)$

$K(\text{combination A}) = K(C_1\text{-}C_2) + K_{series}(C_2\text{-}C_3, C_3\text{-}C_4, C_4\text{-}C_5)$

Combination B:

Group bond $K(C_1\text{-}C_2)$ and bond $K(C_2\text{-}C_3)$ are in series. Meanwhile, group bond $K(C_3\text{-}C_4)$ and bond $K(C_4\text{-}C_5)$ are in series. Then $K_{series}(C_1\text{-}C_2, C_2\text{-}C_3)$ and $K_{series}(C_3\text{-}C_4, C_4\text{-}C_5)$ are in parallel

$K_{series}(C_1\text{-}C_2, C_2\text{-}C_3) = [K(C_1\text{-}C_2)^{-1} + (K(C_2\text{-}C_3)\cos(\text{beta}))^{-1}]^{-1}$

$K_{series}(C_3\text{-}C_4, C_4\text{-}C_5) = [K(C_3\text{-}C_4)^{-1} + (K(C_4\text{-}C_5)\cos(\text{beta}))^{-1}]^{-1}$

$K_{series}(C_1\text{-}C_2, C_2\text{-}C_3)$ and $K_{series}(C_3\text{-}C_4, C_4\text{-}C_5)$ are in parallel

Hence, $K(\text{combination B}) = K_{series}(C_1\text{-}C_2, C_2\text{-}C_3) + K_{series}(C_3\text{-}C_4, C_4\text{-}C_5)$

Combination C:

Group bond $K(C_1\text{-}C_2)$, bond $K(C_2\text{-}C_3)$ and bond $K(C_3\text{-}C_4)$ in series. Then $K_{series}(C_1\text{-}C_2, C_2\text{-}C_3, C_3\text{-}C_4)$ and $K(C_4\text{-}C_5)$ in parallel

$K_{series}(C_1\text{-}C_2, C_2\text{-}C_3, C_3\text{-}C_4) = [K(C_1\text{-}C_2)^{-1} + (K(C_2\text{-}C_3)\cos(\text{beta}))^{-1} + K(C_3\text{-}C_4)^{-1}]^{-1}$

Then $K_{series}(C_1\text{-}C_2, C_2\text{-}C_3, C_3\text{-}C_4)$ and $K(C_4\text{-}C_5)$ are in parallel

Hence, $K(\text{combination C}) = K_{series}(C_1\text{-}C_2, C_2\text{-}C_3, C_3\text{-}C_4) + K(C_4\text{-}C_5)$

The average longitudinal spring constant along the Z axis is

$K_{longitudinal} = (K(\text{combination A}) + K(\text{combination B}) + K(\text{combination C})) / 3$, referring to the structure drawn in Figure S1a

- For the structure in Figure S1a, the lateral spring constant along the Y-axis between atom $C_2$ and $C_3$ is $K(C_2\text{-}C_3)\sin(\text{beta})$. Due to the symmetry along the zigzag path, the lateral spring constant is $K_{lateral} = 2*K(C_2\text{-}C_3)\sin(\text{beta})$ while these two projected covalent bonds are in parallel along the Y-axis

- The tube curvature also triggers a phonon softening as shown in Figure S1b. The average lateral spring constant affected by the tube curvature is $K_{lateral\&curve} = K_{lateral}\cos(\text{gamma})$, where gamma refers to the exterior angle of a polygon. Phonon softening is related to the onset Tc, which indicates that the maximum value of exterior angle is required. However, due to the rotational symmetry along the circumference, the exterior angle remains constant around the circumference.

Based on classical harmonic oscillations, the vibrational frequency of atom $\omega = \text{sqrt}(K_{resultant}/m)$, where m is mass and $K_{resultant} = \text{sqrt}(K_{lateral\&curve}^2 + K_{longitudinal}^2)$.

- A modification of longitudinal spring constant for the (2,1)CNT is also needed because the $C_1$, $C_2$, $C_3$, $C_4$ and $C_5$ (yellow balls) are not within a 2D plane (Figure S1d), which means that the longitudinal vibration is also affected by the tube curvature (i.e. $K_{longitudinal\&curve}$). The longitudinal spring constant is first calculated without considering the chiral angle. This value is then multiplied by cos(alpha), where alpha ~ 70 degrees represents the maximum change in the trajectory (brown line) for the longitudinal components. Hence, $K_{resultant}$ = sqrt($K_{lateral\&curve}^2$+ $K_{longitudinal\&curve}^2$).
- The coulomb pseudopotential as a function of Thomas vector, DOS($E_F$), electron concentration, and electron phonon coupling and Debye temperature, etc [30] is ~0.14 for the ultrathin CNT in this work, which can be approximately it to zero. Since the electron-phonon coupling is already much larger than the Coulomb pseudopotential for the 100 K superconductor [21], this approximation becomes more accurate for room-temperature superconductors
- Lattice distortion may exist under doping. The overall tube curvature is analyzed by using the average tube diameter, where a few percent increase in the average CNT diameter is observed under doping.

Section 2: Tc calculations (Example)

$\lambda = \langle g_{kk}^2 \rangle$ DOS($E_F$) / (M $\langle \omega_{softening}^2 \rangle$ is used. [21]

Based on the scale factor approach [21], $T_c^P = T_c^Q \left| \frac{\lambda_P - \mu_P}{\lambda_Q - \mu_Q} \frac{1}{f(T_D)} \right|$

Since $\mu \ll \lambda \Rightarrow \mu \sim 0$ and $f(T_D) \sim 1$, the ratio of Tc ~ the ratio of $\lambda$

$\langle \omega_{softening}^2 \rangle$ ratio ~ $\langle \omega^2 \rangle$ ratio x $K_{resultant}$ ratio, where $\langle \omega^2 \rangle$ can be found by ab-initio input.

- (3,0)CNT

| | $\langle g_{kk}^2 \rangle$ ratio | DOS($E_F$) ratio | $\langle \omega^2 \rangle$ ratio | $K_{resultant}$ ratio | $\langle \omega_{softening}^2 \rangle$ ratio | M Ratio |
|---|---|---|---|---|---|---|
| (3,0)CNT | 5.64 | 1.78 | 0.99 | 0.74 | 0.73 | 1 |
| (5,0)CNT reference | 1 | 1 | 1 | 1 | 1 | 1 |

The experimental Tc of (5,0)CNT is 7.6K [29], the scale factor approach gives the calculated Tc of (3,0)CNT is 103K

$T_c$ = 7.6*($\langle g_{kk}^2 \rangle$ ratio) * (DOS($E_F$) ratio) / (M ratio * $\langle \omega_{softening}^2 \rangle$ ratio)

= 7.6*5.64*1.78/(1*0.73) = 104K

- (2,1)CNT

| | $<g_{kk}^2>$ ratio | $DOS(E_F)$ ratio | $<\omega^2>$ ratio | $K_{resultant}$ ratio | $<\omega_{softening}^2>$ ratio | M Ratio |
|---|---|---|---|---|---|---|
| (2,1)CNT | 3.76 | 1.34 | 0.84 | 0.34 | 0.28 | 1 |
| (5,0)CNT reference | 1 | 1 | 1 | 1 | 1 | 1 |

The experimental Tc of (5,0)CNT is 7.6K [29], the scale factor approach gives the calculated Tc of (2,1)CNT is 137K

$T_c = 7.6*(<g_{kk}^2>\text{ ratio}) * (DOS(E_F)\text{ ratio}) / (M\text{ ratio} * <\omega_{softening}^2>\text{ ratio})$

$= 7.6*3.76*1.34/(1*0.28) = 137K$

- 3.5% B-doped (2,1)CNT (relaxed shape)

| | $<g_{kk}^2>$ ratio | $DOS(E_F)$ ratio | $<\omega^2>$ ratio | $K_{resultant}$ ratio | $<\omega_{softening}^2>$ ratio | M ratio |
|---|---|---|---|---|---|---|
| (2,1)CNT+3.5%B | 6.08 | 1.80 | 0.88 | 0.35 | 0.30 | 1 |
| (5,0)CNT reference | 1 | 1 | 1 | 1 | 1 | 1 |

The experimental Tc of (5,0)CNT is 7.6K [29], the scale factor approach gives the calculated Tc of 3.5% B-doped (2,1)CNT is 277K

$T_c = 7.6*(<g_{kk}^2>\text{ ratio}) * (DOS(E_F)\text{ ratio}) / (M\text{ ratio} * <\omega_{softening}^2>\text{ ratio})$

$= 7.6*6.08*1.80 /(1*0.30) = 277K$